\documentclass[conference]{IEEEtran}

\usepackage{cite}
\usepackage{amsmath,amssymb,amsfonts}
\usepackage{booktabs}
\usepackage{array}
\usepackage{graphicx}
\usepackage{tikz}
\usepackage{textcomp}
\usepackage{xcolor}
\usepackage{url}
\usepackage[hidelinks]{hyperref}
\usetikzlibrary{arrows.meta,backgrounds,calc,fit,positioning}

\definecolor{hkvmblue}{HTML}{0072B2}
\definecolor{hkvmorange}{HTML}{E69F00}
\definecolor{hkvmgreen}{HTML}{009E73}
\definecolor{hkvmvermillion}{HTML}{D55E00}
\definecolor{hkvmgray}{HTML}{666666}

\def\BibTeX{{\rm B\kern-.05em{\sc i\kern-.025em b}\kern-.08em
    T\kern-.1667em\lower.7ex\hbox{E}\kern-.125emX}}

\newcommand{\method}{HKVM-RAG}
\newcommand{\whg}{Weighted-HG-KV}
\newcommand{\kgppr}{KG-PPR}

\IEEEoverridecommandlockouts

\begin{document}

\title{\method: Key-Value-Separated Hypergraph Evidence Organization for Multi-Hop RAG}

% ICDE 2027 uses single-blind review. Replace this placeholder with the final
% author and affiliation block before submission; ORCIDs are entered in CMT.
\author{
    \parbox{\dimexpr\linewidth-2em}{
        \centering
        Mingyu Zhang\textsuperscript{1}, Fanghui Sun\textsuperscript{1}, Chunjing Xiao\textsuperscript{2}, Ying Ma\textsuperscript{1,*} \\
        \textsuperscript{1}Faculty of Computing, Harbin Institute of Technology, Harbin, China \\
        \textsuperscript{2}School of Computer and Information Engineering, Henan University, Kaifeng, China \\
        25b903142@stu.hit.edu.cn, Sunfanghui@hit.edu.cn, chunjingxiao@gmail.com, y.ma@hit.edu.cn
    }%
    \thanks{\textsuperscript{*}Corresponding author.}
}

\maketitle

\begin{abstract}
Multi-hop RAG retrieval must find relevant passages and organize them into evidence units that expose answer-supporting chains. We study this layer as a controlled data-engineering problem, testing whether a retrieval key space built from answer-path hyperedges improves passage organization over pairwise graph keys under a fixed extraction substrate and matched retrieval budget. We present HKVM-RAG, a key-value-separated evidence index that constructs answer-path hyperedges from cached LLM evidence tuples as retrieval keys while keeping passage text as values. It has two bounded roles in this protocol. As a standalone structured retriever, weighted hypergraph key-value retrieval improves over pairwise KG-PPR on 2WikiMultiHopQA and MuSiQue, while HotpotQA marks the boundary where structure alone is insufficient. As an evidence-control signal, a lightweight controller over frozen ColBERTv2 and WHG-KV features improves over ColBERTv2 by 11.084, 6.763, and 5.966 F1 points on 2WikiMultiHopQA, MuSiQue, and HotpotQA; source-level ablations show that the gain is WHG-KV-specific rather than a generic effect of adding any structured source. Transfer and target-calibration audits show that the dense/HKVM rank-score signal transfers across benchmarks under frozen prediction artifacts, while target-domain calibration remains useful and sample-sensitive. The bounded conclusion is that key-value-separated hypergraph organization complements lexical and dense retrievers as a reusable evidence-control layer rather than replacing them.\footnotemark
\end{abstract}
\footnotetext{Code and data are available at: \url{https://github.com/Mingyu-Zh/HKVM-RAG}.}


\section{Introduction}

We study multi-hop RAG as a data-engineering problem: how to organize evidence under fixed retrieval budgets. Retrieval-augmented generation grounds model outputs in external evidence by coupling parametric language models with retrieved non-parametric knowledge \cite{lewis2020rag,guu2020realm,izacard2020fid}. Recent database-community discussions frame RAG as a data-management and system-design problem involving retrieval, indexing, and evaluation infrastructure \cite{fan2025ragdatamgmt,khan2025ragdm}. In this paper, the system question is how passages, vector scores, extracted relations, and reproducible budgets are arranged before answer generation. We therefore treat candidate passages, extracted tuples, structured keys, prediction tables, and manifests as auditable data artifacts in an evidence-indexing workload. Recent RAG-memory systems extend this idea with explicit retrieval structures for questions whose evidence is distributed across multiple passages \cite{hipporag,hipporag2,ecphoryrag}. In multi-hop question answering benchmarks \cite{ho2020_2wikimultihop,trivedi2022musique,yang2018hotpotqa}, the retriever must do more than find individually relevant passages. It must organize passages into evidence units that preserve how facts connect across hops. A system can fail even when the answer-bearing text is present in the corpus if the retrieved context does not expose the evidence chain needed by the reader.

\begin{figure}[!t]
\centering
\includegraphics[width=\columnwidth]{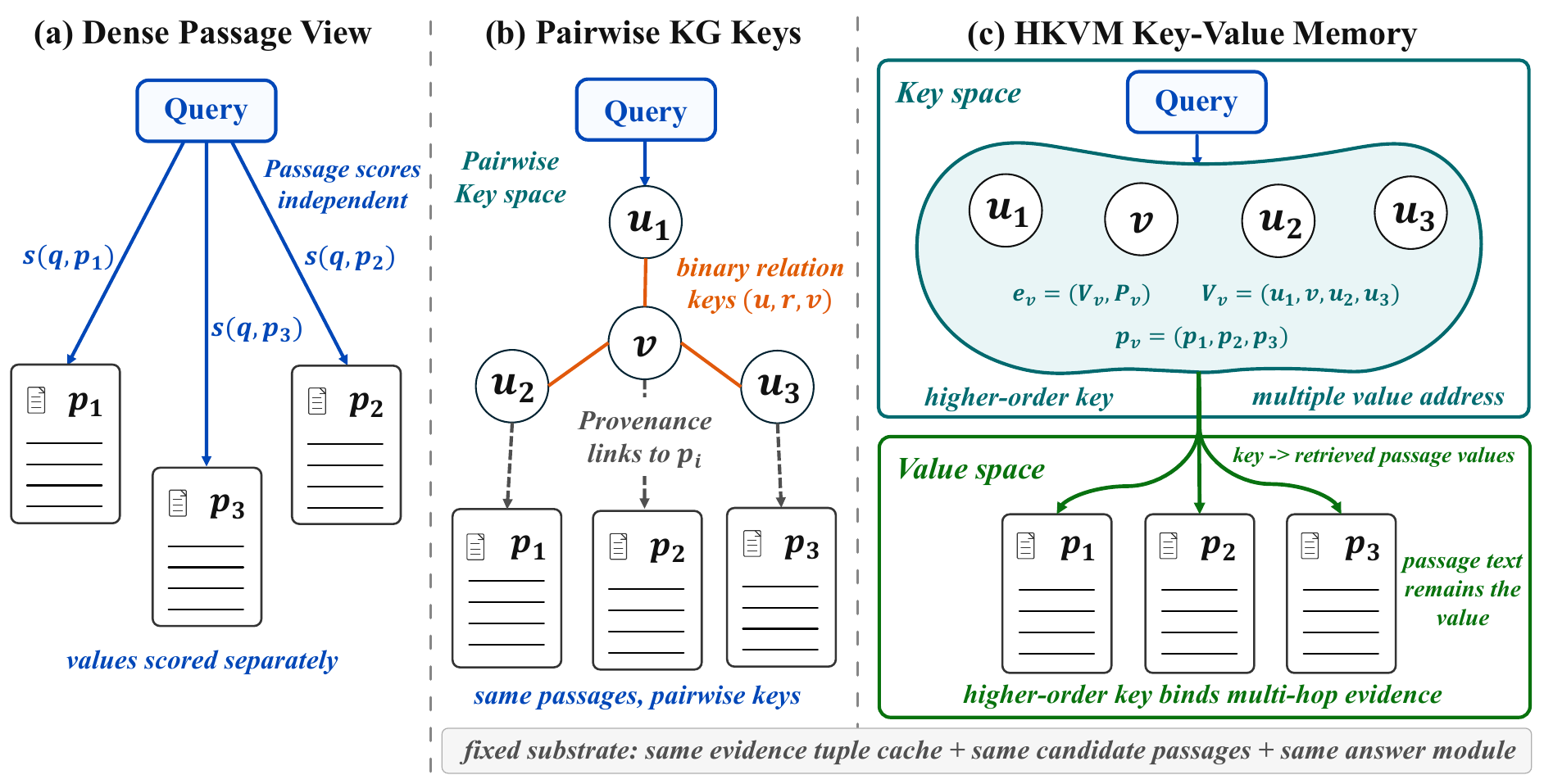}
\caption{Conceptual comparison of three retrieval views under a fixed evidence substrate. (a) Dense retrieval scores passage values independently. (b) Pairwise KG retrieval uses binary entity-relation keys over entity vertices and maps them back to passage values through provenance links. (c) \method{} separates key-side evidence organization from value-side passage text: an answer-path hyperedge key connects multiple entity vertices and maps back to its associated passage-value addresses. The schematic illustrates the mechanism tested in the cached-extraction setting.}
\label{fig:concept_hkvm_kv}
\end{figure}

Two major retrieval routes address different parts of this problem. Lexical, dense, and late-interaction retrievers are strong matching models \cite{bm25,karpukhin2020dpr,contriever,colbertv2}, but their scores do not explicitly encode the structured evidence units used by a multi-hop answer. Graph-based RAG and knowledge-graph memory systems make associations explicit by indexing extracted entities and relations and traversing the resulting graph \cite{edge2024graphrag,sun2024tog,hipporag,hipporag2,ecphoryrag}. This structure is useful, but the retrieval keys remain primarily pairwise or entity-centered graph objects. When answer evidence is better represented as a group of connected facts, flattening that group into independent edges can weaken the retrieval signal.

Hypergraph learning methods provide a natural representation for higher-order relations \cite{hgnn,yadati2019hypergcn,chien2022allset,antelmi2023hypergraphsurvey}, and recent RAG systems have begun to explore hypergraph-driven retrieval and memory \cite{hyperrag,hgmem}. The remaining systems question is how to test such higher-order retrieval keys without also changing the underlying evidence substrate. Improvements can become ambiguous when multiple experimental axes change together. Extraction design and model scale are two major confounds. Online filtering budgets and high-order relation schemas add further confounds. The result may reflect a stronger extraction pipeline rather than the key structure itself. This paper isolates the key-space question: given the same evidence tuple cache and candidate passages, does an answer-path hypergraph key space organize multi-hop evidence better than a pairwise graph key space?

We propose \method, a structured evidence-indexing design. HKVM is an operational key-value memory abstraction: extracted entity vertices, pairwise relation edges, answer-path hyperedge keys, and confidence weights form the retrieval-key side, while passage text remains the value side used for answer extraction. The system constructs answer-path hyperedges from LLM-extracted evidence tuples and uses them as retrieval keys over passage values. The paper addresses two connected questions. First, do hypergraph keys improve evidence organization over pairwise graph keys under a fixed extraction substrate? Second, are the resulting hypergraph key scores useful as control signals alongside strong lexical and dense passage scores? Figure~\ref{fig:concept_hkvm_kv} contrasts the three retrieval views: dense scoring over separate passage values, query-seeded pairwise KG keys with provenance links to passages, and a single answer-path hyperedge key that maps back to multiple passage values.

We evaluate this mechanism rather than a leaderboard claim. Under a shared tuple cache, \whg{} improves over a HippoRAG-style \kgppr{} reference by +3.426 F1 on 2WikiMultiHopQA and +3.592 F1 on MuSiQue, while HotpotQA supplies the boundary case where standalone structure is not a dense-retrieval replacement. Oracle answer-path selection raises 2WikiMultiHopQA F1 from 79.299 to 90.884 and MuSiQue F1 from 39.925 to 74.330, pointing to support selection as a major bottleneck. Treating WHG-KV as a control signal, a dense-aware controller reaches 88.846 F1 on 2WikiMultiHopQA, 65.073 F1 on MuSiQue, and 85.810 F1 on HotpotQA. Source-level diagnostics show that matched non-WHG sources do not match these gains, while transfer and target-calibration audits show that the dense/HKVM rank-score signal transfers across the tested benchmarks under frozen prediction artifacts, with remaining target-domain and sample-composition sensitivity. The resulting claim is deliberately narrow: \method{} is a lightweight evidence-organization component whose key-side signal can complement lexical and dense retrievers under a controlled protocol.

The contributions are:
\begin{itemize}
    \item We formulate multi-hop RAG retrieval as fixed-budget evidence organization and introduce answer-path hyperedge keys that assemble cross-passage evidence from cached LLM tuples while keeping passage text as the value side for answer extraction.
    \item We define a controlled comparison protocol that isolates key-space design by holding the evidence tuple cache, candidate passages, reader, and evaluation budget fixed across pairwise graph and hypergraph rows.
    \item We diagnose the standalone hypergraph boundary with oracle gap decomposition and train-to-dev support scoring, identifying support selection and ranking as major repairable bottlenecks under the current cached extraction substrate.
    \item We introduce a dense-aware HKVM controller and show that WHG-KV rank/score signals complement ColBERTv2, BM25, and Contriever under the tested protocol; source-level diagnostics show that matched non-WHG structured sources do not reproduce the gain.
    \item We audit transfer and target calibration over frozen prediction artifacts, narrowing the safe mechanism claim to dense/HKVM rank-score calibration and ruling out target-data independence or monotonic sample-efficiency claims.
\end{itemize}

We release a reproducibility-oriented artifact that includes code, processing scripts, frozen manifests, prediction files, and redistributable processed outputs needed to verify the paper-facing tables and figures.

\section{Related Work}

\subsection{Lexical, Dense, and Multi-Hop Retrieval}

We organize related work by the data state each line contributes to our controlled evidence-indexing view: passage scoring, graph-key memory, and hypergraph key-value memory. Retrieval-augmented generation and retrieval-augmented language-model pre-training show that external evidence can improve knowledge-intensive NLP and open-domain question answering \cite{lewis2020rag,guu2020realm}. Retrieval-reader architectures such as FiD fuse independently encoded retrieved passages during generation \cite{izacard2020fid}. At the retriever level, BM25 remains a robust lexical reference point \cite{bm25}, dense passage retrieval learns query and passage representations in a shared vector space \cite{karpukhin2020dpr}, Contriever represents unsupervised dense retrieval \cite{contriever}, and ColBERTv2 uses late interaction to retain token-level matching signals \cite{colbertv2}. At the systems layer, dense retrieval depends on approximate nearest-neighbor and vector-data infrastructure such as HNSW, FAISS, anisotropic vector quantization, and vector database systems \cite{malkov2020hnsw,johnson2021faiss,guo2020scann,wang2021milvus}. Multi-hop retrieval systems such as MDR, Baleen, and IRCoT make evidence gathering iterative or reasoning-guided \cite{xiong2021mdr,khattab2021baleen,trivedi2023ircot}. We include these systems as strong matching and multi-hop retrieval context. Our work complements them by isolating a different variable: whether changing the structured key space — from pairwise KG edges to answer-path hyperedges — improves passage organization when the evidence substrate and answer budget are held fixed.

\subsection{Knowledge-Graph Memory for RAG}

Graph-based RAG and KG-augmented reasoning methods organize evidence through entity-relation graphs or KG traversal \cite{edge2024graphrag,sun2024tog}. This connects RAG evidence retrieval to a broader data-management setting in which querying knowledge graphs, including graph traversal and path-oriented access, is a long-standing problem \cite{khan2023kgquery}. Knowledge-graph memory systems make associations explicit by indexing extracted entities and relations and then traversing the resulting graph. HippoRAG organizes LLM-extracted triples into a KG and uses personalized PageRank for associative retrieval \cite{hipporag}. HippoRAG 2 strengthens this line with passage nodes, query-to-triple linking, and online LLM triple filtering \cite{hipporag2}. EcphoryRAG follows a related entity-centric route, where cue entities activate entity-centered engrams for associative search \cite{ecphoryrag}. These systems support the premise that explicit structure can help multi-hop retrieval, but their retrieval keys remain pairwise or entity-centered graph units, often coupled with additional online filtering.

HippoRAG 2 and EcphoryRAG are the closest contemporaneous KG-memory systems to our setting \cite{hipporag2,ecphoryrag}. We position them as full-system KG-memory references, while our numeric matrix uses a controlled key-space isolation protocol. HippoRAG 2 uses Llama-3.3-70B-Instruct, NV-Embed-v2, and online LLM triple filtering \cite{hipporag2}; EcphoryRAG reports Phi4/Ollama, bge-m3 embeddings, 500 random questions per dataset, and dataset-specific context tuning \cite{ecphoryrag}. These differences in models, splits, prompts, filtering policies, and context budgets would change several axes at once, so the main comparison isolates pairwise KG keys and answer-path hypergraph keys under a shared substrate.

\subsection{Hypergraph Retrieval and Key-Value Memory}

Hypergraph neural networks and representation-learning methods model higher-order relations that are not captured cleanly by independent pairwise edges \cite{hgnn,yadati2019hypergcn,chien2022allset,antelmi2023hypergraphsurvey}. Hyper-RAG and HGMem bring related hypergraph ideas into RAG and working memory \cite{hyperrag,hgmem}. Our construction is narrower: answer-path hyperedges are assembled from the same evidence tuple cache used by the pairwise \kgppr{} baseline, so the experiment tests whether changing the retrieval key space improves evidence organization under a fixed extraction substrate. We include Co-occurrence-HG as a Hyper-RAG-style matched source constructed from the same cached evidence tuples and candidate passages. This row tests whether a generic co-occurrence hypergraph source is sufficient under the same controller protocol, while avoiding additional extraction, high-order schema, filtering, retrieval-pipeline, and answer-assembly differences.

The memory framing is also scoped. End-to-end and key-value memory networks support multi-hop reads over external memory \cite{sukhbaatar2015memn2n,miller2016kvmemn2n}, Modern Hopfield networks connect associative memory and attention \cite{modernhopfield}, and cognitive key-value memory motivates separating retrieval keys from stored values \cite{kvbrain}. \method{} adopts this separation operationally: graph or hypergraph objects form the traversal key side, while passages remain the value side used for answer extraction. We use these analogies to motivate retrieval organization, not to claim biological fidelity.

\section{Problem Formulation}

Given a query $q$ and a candidate passage set $P=\{p_i\}$, the evidence-organization layer must rank passages and produce an answer context. We study this layer after candidate passage construction, without claiming a new full-corpus indexing stack. This makes key-space design a controlled intervention over fixed candidate passages, extracted evidence, and answer modules. A dense retriever assigns direct scores $s(q,p_i)$ to candidate passages, treating them primarily as separate values. Each passage $p_i$ is also processed by an LLM extractor to produce evidence tuples $R_i=\{(h,r,t,p_i,c_f,c_s,c_b)\}$. Let $R=\bigcup_i R_i$ and let $V$ be the set of entity vertices induced by tuple heads and tails. A pairwise graph baseline builds $G=(V,E_g)$ from the relation triples in $R$ and retrieves by graph traversal over binary relation keys. \method{} builds a hypergraph $H=(V,E_h)$ over the same entity vertices and extracted tuple cache, where each hyperedge key connects a group of entity vertices and passage-value addresses that participate in a candidate answer path.

We distinguish memory keys from memory values. Entity vertices, pairwise relation edges, hyperedge keys, and their weights define the key space used for retrieval. Passage text remains the value side used for answer support. A pairwise key links two entities at a time, whereas an answer-path hyperedge key can bind several entities and then map score mass back to its associated passage values. All structured methods finally produce passage scores $s(p_i)$ and pass the top-ranked passages to the same answer module. The central comparison is not between different answer modules, but between two structured key spaces over the same extracted evidence: pairwise \kgppr{} and answer-path hypergraph traversal.

\section{Method}

\begin{figure*}[!t]
\centering
\includegraphics[width=\textwidth]{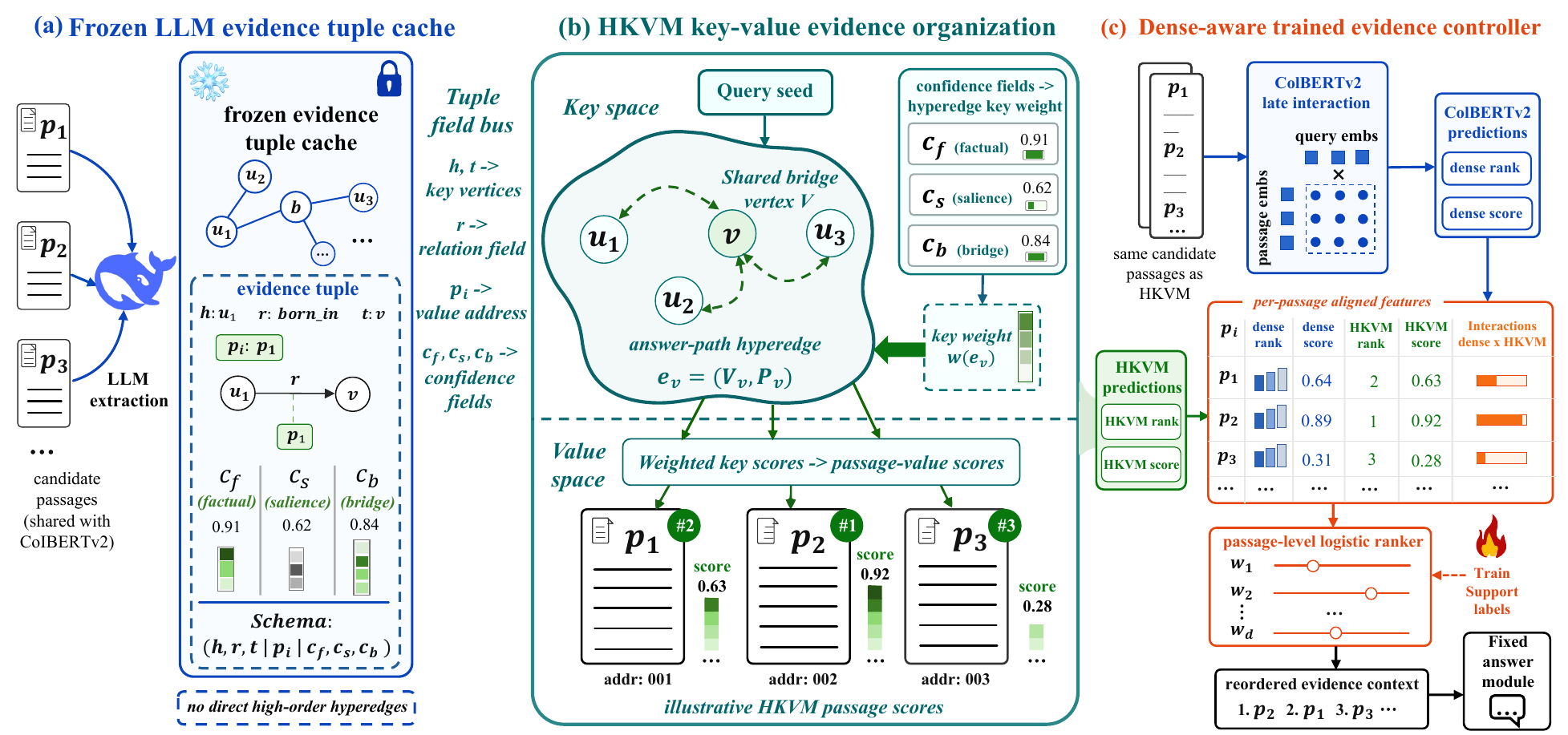}
\caption{\method{} data states and dense-aware evidence control. (a) A frozen LLM evidence tuple cache stores passage-level relation triples, passage ids, and confidence fields $(c_f,c_s,c_b)$; tuple heads and tails define entity vertices, while relation phrases remain relation fields. (b) HKVM assembles answer-path hyperedge keys from shared bridge vertices, weights them with extractor confidence, diffuses query-seeded key scores, and projects the result back to passage-value scores through passage addresses. (c) The dense-aware controller aligns frozen ColBERTv2 and HKVM rank/score predictions by passage id and trains a passage-level logistic ranker on train-side support labels to reorder the final evidence context for the fixed answer module.}
\label{fig:concept_dense_controller}
\end{figure*}

\subsection{Overview}

\method{} is a structured evidence-organization procedure over a fixed candidate-passage substrate. The comparison unit is a query-specific evidence table containing candidate passage ids, dense scores, extracted tuple fields, structured scores, and final answer-module inputs. Figure~\ref{fig:concept_dense_controller} tracks these data states rather than a separate end-to-end retrieval pipeline. An LLM first creates a seed-invariant passage-level tuple cache with local triples, passage ids, and confidence fields $(h,r,t,p_i,c_f,c_s,c_b)$; it does not emit high-order hyperedges. HKVM then turns heads and tails into entity vertices, relation phrases into pairwise fields, shared bridge vertices into answer-path hyperedge keys, and passage ids into value addresses. Confidence fields weight key-side objects before query-seeded diffusion maps key scores back to passage values. In the dense-aware setting, frozen ColBERTv2 and HKVM predictions are aligned by passage id into rank/score features, and a passage-level controller reorders the final evidence context. Table~\ref{tab:main_results} tests the key-space intervention; the controller setting tests whether the same hypergraph signal is more useful as a dense complement than as a standalone retriever.

\subsection{Shared LLM Evidence Tuple Extraction}

All structured methods start from a shared passage-level evidence tuple cache. The relation component is triple-only: the extractor emits local $(h,r,t)$ relation triples with passage ids and confidence fields, but does not emit direct high-order hyperedges. For each passage $p_i$, the extractor returns a small set of evidence tuples
\begin{equation}
(h,r,t,p_i,c_f,c_s,c_b),
\end{equation}
where $h$ and $t$ are entity strings, $r$ is a relation phrase, and $c_f$, $c_s$, and $c_b$ denote factual confidence, semantic salience, and bridge potential. The extraction schema is passage-level and does not ask the LLM to produce high-order hyperedges directly. This keeps the LLM output local and auditable, while leaving cross-passage grouping to the deterministic answer-path assembly step.

Evidence tuple extraction is cached once and treated as seed-invariant. The three evaluation seeds reuse the same tuples and do not trigger new LLM calls, so structured rows differ in key-space construction and traversal rather than stochastic extraction.

\subsection{Answer-Path Hyperedge Assembly}

Pairwise graph memory connects evidence units through binary relations, which can fragment a multi-hop evidence group into separate edges. \method{} constructs answer-path hyperedges from the shared evidence tuples by grouping local triple components that share a bridge vertex. The notation below shows the entity, relation, and passage components; the associated confidence fields are carried with the grouped tuples and later aggregated onto the hyperedge. For a bridge vertex $v$, define
\begin{equation}
G_v=\{(h,r,t,p,c_f,c_s,c_b)\in R: v=h \ \text{or}\ v=t\}.
\end{equation}
If $|G_v|\geq 2$ and the union of heads and tails contains at least three distinct entity vertices, \method{} creates an answer-path hyperedge key
\begin{equation}
e_v=\left(V_v,P_v\right),
\end{equation}
where $V_v=\bigcup_{(h,r,t,p,c_f,c_s,c_b)\in G_v}\{h,t\}$ is the grouped entity-vertex set and $P_v=\{p:(h,r,t,p,c_f,c_s,c_b)\in G_v\}$ is the associated passage-address set. The hyperedge key stores the connected entity vertices, the passage ids that can receive score mass, and aggregated extractor fields inherited from the grouped tuples.

The final default setting constructs these hyperedges without dev-gold support and without direct LLM high-order relations. Direct high-order extraction was kept out of the default because the diagnostic cache showed passage-local relations rather than cross-passage evidence paths. Answer-path assembly therefore provides the high-order key structure while preserving a local, auditable LLM evidence-tuple substrate.

\subsection{Weighted Hypergraph Key Traversal}

\whg{} scores hypergraph keys with extractor-derived confidence signals. For a hyperedge key $e\in E_h$, the default structured setting computes an LLM-confidence score
\begin{equation}
a(e)=\frac{\lambda_f c_f(e)+\lambda_s c_s(e)+\lambda_b c_b(e)}
{\lambda_f+\lambda_s+\lambda_b},
\end{equation}
where $c_f(e)$, $c_s(e)$, and $c_b(e)$ are inherited from the grouped tuples. The default uses $\lambda_f=0.20$, $\lambda_s=0.40$, and $\lambda_b=0.40$. The hyperedge weight is then
\begin{equation}
w(e)=1+(\beta-1)a(e),
\end{equation}
with $\beta=2.0$. Thus all three extractor confidence fields contribute to the key weight, with semantic salience and bridge potential weighted higher than factual confidence. Weighted-KG-PPR uses the analogous confidence weighting on pairwise graph edges, which lets the experiments separate the effect of confidence weighting from the effect of changing the key structure.

Retrieval starts by seeding vertices from query-entity matches and token overlap. Hypergraph diffusion alternates between vertex scores and hyperedge scores. At each step, a hyperedge receives the weighted average score of its incident vertices, and vertices receive normalized mass back from incident hyperedges. With diffusion parameter $\rho=0.35$, one update can be written as
\begin{equation}
x_v^{(t+1)}=\rho x_v^{(0)}+(1-\rho)\sum_{e\ni v}
\frac{w(e)}{d(v)|e|}\sum_{u\in e}x_u^{(t)},
\end{equation}
where $d(v)$ is the number of incident hyperedges. The default system uses traversal depth $T=1$. Additional diffusion depth $T=2$ is reported as an ablation rather than as the default system.

After diffusion, passage scores combine two sources: vertex evidence from evidence tuples and direct hyperedge-to-passage evidence from $P_v$. This maps key-side evidence organization back to passage values, producing HKVM passage ranks and scores before the common answer module or the dense-aware controller is applied.

\subsection{Pairwise Graph Reference}

\kgppr{} is the low-order structured reference. It uses the same evidence tuple cache and candidate-passage budget, but restricts retrieval structure to pairwise graph traversal over the local triple components. Each tuple contributes an undirected edge between its head and tail. Query seeds are constructed in the same way as for \whg{}, and personalized PageRank is run with restart parameter $\alpha=0.5$, 40 iterations, and tolerance $10^{-7}$. Passage scores are obtained by summing the final scores of the tuple endpoints that occur in each passage.

This reference isolates the effect of replacing pairwise keys with answer-path hypergraph keys. Static-HG removes confidence weighting from the same answer-path hypergraph, while Weighted-KG-PPR adds the same LLM-confidence weighting to pairwise graph edges. These control rows make the method comparison about key structure and weighting rather than about extraction, passages, or answer-module changes.

\subsection{Dense-Aware Evidence Controller}

The standalone structured comparison shows whether hypergraph keys improve over pairwise graph keys, but it does not assume that hypergraph retrieval should replace dense retrieval. The dense-aware controller instead treats ColBERTv2 and HKVM outputs as two frozen prediction bundles over the same candidate passage ids. For each query-passage row, we align the dense and HKVM predictions by passage id and form a per-passage feature vector containing ColBERTv2 rank and score, HKVM rank and score, reciprocal-rank features, normalized score features, and dense-HKVM interaction terms. These are inference-time ranking signals; the controller does not use dev labels at inference time and does not run a new dense encoder or a new LLM extraction pass.

We use three names for the train-to-dev ranking stages. The train-derived support scorer is the first HKVM-only passage ranker over cached structured outputs. The learned HKVM calibration scorer is the stronger HKVM-only variant over candidate-evidence features. The dense-aware controller is the final passage-level ranker that adds dense rank/score features to HKVM signals. All stages train on train-side labels and evaluate on frozen development predictions.

The controller is a passage-level logistic ranker trained on train-side support labels. To avoid train-side HKVM stacking artifacts, the final paper-facing run uses out-of-fold HKVM predictions for the controller's train-side HKVM inputs. Each train example is predicted only by a model trained on the other folds, while dev evaluation uses the fixed learned HKVM calibration predictions. The final configuration uses 80 epochs, learning rate 0.01, $L_2=10^{-4}$, negative ratio 8, top-10 candidate retrieval, and top-5 answer context. We report dense-only, HKVM-only, and dense+HKVM controller variants to separate source-only evidence and learned passage-level control.

For the source-level controller ablation, the first-stage prediction source, train-to-dev split, controller family, and hyperparameters are fixed while the structured bundle is swapped. The primary ablation uses ColBERTv2 as the first-stage source and replaces only the structured prediction bundle with matched Co-occurrence-HG, Static-HG, Weighted-KG-PPR, or WHG-KV predictions over the same datasets and seeds. We repeat the diagnostic with BM25 and Contriever to test whether the WHG-KV ordering is tied to late-interaction ColBERTv2. Non-WHG train-side files are generated from existing cached train extraction artifacts with no new LLM calls, and dev-side files reuse the frozen structured outputs from the main result matrix.

\section{Experimental Setup}

\subsection{Evaluation Splits and Metrics}

The experimental setup fixes data states before comparing evidence-organization choices, without ranking unrestricted retrieval systems. We evaluate on three public multi-hop QA benchmarks: 2WikiMultiHopQA \cite{ho2020_2wikimultihop}, MuSiQue \cite{trivedi2022musique}, and HotpotQA \cite{yang2018hotpotqa}. This is the same family of benchmarks used by HippoRAG \cite{hipporag}, but we report our exact official development splits. Table~\ref{tab:eval_splits} summarizes the exact splits: the full 2WikiMultiHopQA development set, the MuSiQue development answerable subset, and the HotpotQA development distractor set. No hidden-test, leaderboard, or server-submission results are included in the current result matrix; all claims are development-split, frozen-artifact evidence-organization claims.

\begin{table}[t]
\caption{Evaluation splits used in the main result matrix.}
\label{tab:eval_splits}
\centering
\small
\begin{tabular}{llr}
\toprule
Dataset & Split & Examples \\
\midrule
2WikiMultiHopQA & Dev full & 12576 \\
MuSiQue & Dev answerable & 2417 \\
HotpotQA & Dev distractor & 7405 \\
\bottomrule
\end{tabular}
\end{table}

Metrics are all-recall at 10 (AR@10), answer F1, and exact match (EM). AR@10 measures whether all gold supporting passages are present in the top-10 retrieved passages. F1 and EM are computed from the common extractive answer output. The fixed-substrate key-space comparison uses seed-13 unique-example paired bootstrap with 1,000 samples and $\alpha=0.05$. Train-to-dev controller and source-level diagnostics use 5,000-sample paper-export paired bootstrap over frozen prediction files; the supplementary material reports the corresponding provenance and intervals by export family.

\subsection{Matched-Budget Protocol}

All methods share the same passage preprocessing and answer-module budget: 1,200-token chunks with 100-token overlap and top-10 retrieval. The extractive answer module receives the top five passages under a 3,000-token context budget. This setup evaluates evidence indexing and passage ordering within a controlled candidate-passage substrate; it is not intended to measure deployment-time corpus indexing throughput or full-corpus retrieval recall. The answer module and evaluation script are fixed across rows, so the main comparison concerns retrieval and memory organization rather than answer-module variation.

\subsection{Compared Methods}

The main table includes BM25, Contriever, ColBERTv2, \kgppr{}, Weighted-KG-PPR, Co-occurrence-HG, Static-HG, and \whg{}. This covers lexical retrieval, dense retrieval, pairwise structured memory, matched co-occurrence hypergraph memory, unweighted answer-path hypergraph memory, and weighted hypergraph key memory within the same candidate-passage budget. ColBERTv2 is implemented as candidate-passage late interaction over the same candidate chunks, not as a separate corpus-level ColBERT index. The Co-occurrence-HG row is a Hyper-RAG-style matched baseline constructed from the same shared evidence tuple cache and candidate passages; it is included to control for generic co-occurrence hypergraph structure, not to reproduce the original Hyper-RAG pipeline.

\subsection{Fairness and Reproducibility}

Dense and structured rows were merged from compatible source runs because of PyTorch/CUDA compatibility constraints. The merge happens at the prediction-table level: each row preserves the query id, passage id, method score, seed, split, and evaluation setting recorded in the source manifest. The comparison preserves identical data splits, candidate-passage budgets, evaluation metrics, answer-module settings, and seed set (13, 29, and 43). This controlled-workload design follows the same broad benchmarking principle used in database evaluation: fixed workloads and explicit evaluation states make system comparisons more interpretable \cite{erling2015ldbc}. The merged manifest records the source split and has no missing rows. This separation ensures that dense encoding does not share GPU memory or runtime state with structured key-space construction, so the measured performance of each method is independently attributable.

\subsection{LLM Extraction Disclosure}

Structured methods use DeepSeek V4 Flash with a deterministic relation-triple extraction schema: no direct high-order extraction, temperature 0.0, passage-level extraction units, and at most five evidence tuples per passage. Table~\ref{tab:llm_cache} reports the fixed extraction scope; these passage-unit counts are cached extractor inputs, not the top-10 retrieval budget, and they differ because candidate-passage pools differ before the shared retrieval and answer-context budgets are applied. Seeds 13, 29, and 43 reuse the same cached tuples. Train-side extraction scopes are used only for the train-to-dev support scorer, learned HKVM calibration, and dense-aware controller. Gold support is not sent to the extractor; train labels are used only after cache construction. The tuple cache is shared and auditable, but sparser than our earlier heuristic OpenIE cache and not claimed to improve the pairwise KG baseline by itself. Changing the extraction endpoint or model requires a new extraction manifest.

We treat these cached evidence tuples as reproducibility artifacts rather than as new benchmark labels. Released cache files will mark tuple records as AI-generated extraction outputs, preserve source-dataset attributions, and follow the redistribution limits imposed by dataset licenses and provider terms. The artifact plan below specifies the code, manifests, scripts, and hashes used to verify or regenerate the frozen result matrix.

\begin{table}[t]
\caption{LLM extraction scope for fixed-substrate structured evaluation. Units are passage-level cached extractor inputs; detailed manifests and exception logs are provided with the artifact.}
\label{tab:llm_cache}
\centering
\small
\begin{tabular*}{0.9\linewidth}{@{\extracolsep{\fill}} llrr @{}}
\toprule
Use & Dataset split & Examples & Passage units \\
\midrule
Main & 2Wiki dev full & 12576 & 125760 \\
Main & MuSiQue dev ans. & 2417 & 48325 \\
Main & HotpotQA dev dist. & 7405 & 28091 \\
\midrule
Train & 2Wiki train subset & 10000 & 100000 \\
Train & MuSiQue train ans. & 10000 & 199987 \\
Train & HotpotQA train subset & 10000 & 99504 \\
\bottomrule
\end{tabular*}
\end{table}

\subsection{Artifact Availability}

The submission artifact reproduces the paper-facing tables and figures from frozen manifests rather than fresh extraction runs, following database-community emphasis on artifact availability and reproducibility \cite{athanassoulis2022artifacts}. Code, configs, scripts, results, manifests, and supplementary material are hosted on GitHub (\url{https://github.com/Mingyu-Zh/HKVM-RAG}); benchmark data and frozen prediction files are hosted on Hugging Face (\url{https://huggingface.co/datasets/MingY-Zh/HKVM-RAG}). The artifact includes code, processing scripts, extraction schema, prompt templates, manifests, prediction files, random seeds, table and figure scripts, environment notes, and evaluation code. For passages or provider-derived outputs that cannot be redistributed directly, it provides regeneration scripts, hashes, and source-dataset pointers. The supplementary material lists the result families and artifact records used to verify the main tables.

\section{Results}

The results follow the evidence chain implied by the paper's claim. We first isolate the key-space intervention under a shared extraction substrate. We then ask why standalone structure fails in some settings and whether the same signal is more useful as a controller over dense candidates. Finally, we test whether the controller gain is WHG-KV-specific and whether the evidence-control signal transfers beyond target-trained development controllers.

\subsection{Answer-Path Keys Improve Structured Evidence Organization}

To test whether key-space design matters, we first hold extraction, passages, and answer generation fixed. Table~\ref{tab:main_results} reports the complete method matrix. The primary comparison is between the pairwise graph reference, \kgppr{}, and the answer-path hypergraph key memory, \whg{}. These rows share the same evidence tuple cache and differ in the retrieval key structure.

Under this fixed substrate, answer-path hypergraph keys improve over the pairwise graph reference on the two datasets where standalone structure helps answer extraction. On 2WikiMultiHopQA, \whg{} reaches 79.299 F1 and 78.650 EM, compared with 75.873 and 75.358 for \kgppr{}. On MuSiQue, it reaches 39.925 F1 and 39.222 EM, compared with 36.332 and 35.540. Table~\ref{tab:bootstrap} reports paired-bootstrap intervals only for \whg{} minus \kgppr{}; lexical and dense rows in Table~\ref{tab:main_results} are contextual references. The intervals are strictly positive on 2WikiMultiHopQA and MuSiQue, supporting the fixed-substrate key-space claim.

HotpotQA separates evidence coverage from answer utility. In the standalone matrix, \whg{} improves AR@10 over \kgppr{} (98.190 versus 95.381), but its F1 is lower (72.957 versus 73.645), and the paired-bootstrap F1 difference is negative with a confidence interval crossing zero. Denser structured retrieval therefore does not automatically improve answer extraction when lexical and dense baselines are strong; the controller results test whether the same WHG-KV signal can still help learned evidence ordering.

The remaining structured rows separate confidence weighting from key structure. Weighted-KG-PPR adds the same confidence weighting to pairwise KG-PPR and changes F1 by only +0.021 on 2Wiki, +0.046 on MuSiQue, and +0.050 on HotpotQA. Static-HG keeps the answer-path hypergraph key structure without the default weighting; compared with Static-HG, \whg{} is higher by +1.201 and +2.211 F1 on 2Wiki and MuSiQue. This comparison is descriptive; the formal test remains Table~\ref{tab:bootstrap}, and HotpotQA limits the interpretation because \whg{} is 0.383 F1 below Static-HG.

The fixed-substrate comparison shows that key-space design matters, but also exposes limits. On MuSiQue, \whg{} (39.925 F1) remains far below ColBERTv2 (58.309). HotpotQA sharpens the same point: \whg{} improves support coverage over \kgppr{} (AR@10 98.190 vs.\ 95.381) but lowers answer F1. Retrieving more structured evidence is not the same as using it well. We therefore turn from standalone retrieval to evidence control: first by decomposing the selection gap, then by testing whether \whg{} complements dense retrieval rather than replacing it.

\begin{table*}[t]
\caption{Main results under matched candidate-passage budgets. Higher is better for all metrics; bold marks the best F1 and EM within each dataset.}
\label{tab:main_results}
\centering
\small
\begin{tabular*}{1.0\textwidth}{@{\extracolsep{\fill}} l rrr rrr rrr @{}}
\toprule
& \multicolumn{3}{c}{\textbf{2WikiMultiHopQA}} & \multicolumn{3}{c}{\textbf{MuSiQue}} & \multicolumn{3}{c}{\textbf{HotpotQA}} \\
\cmidrule(lr){2-4} \cmidrule(lr){5-7} \cmidrule(lr){8-10}
Method & AR@10 & F1 & EM & AR@10 & F1 & EM & AR@10 & F1 & EM \\
\midrule
BM25           & 99.253 & 71.873 & 71.310 & 41.746 & 46.045 & 45.345 & 99.851 & \textbf{81.835} & \textbf{81.648} \\
Contriever     & 100.000 & 65.498 & 64.711 & 54.199 & 55.565 & 54.903 & 100.000 & 79.930 & 79.730 \\
ColBERTv2      & 100.000 & 77.763 & 77.195 & 49.317 & \textbf{58.309} & \textbf{57.758} & 100.000 & 79.844 & 79.635 \\
\midrule
KG-PPR         & 98.108 & 75.873 & 75.358 & 32.478 & 36.332 & 35.540 & 95.381 & 73.645 & 73.315 \\
Weighted-KG-PPR& 98.108 & 75.894 & 75.374 & 32.727 & 36.378 & 35.581 & 95.381 & 73.695 & 73.369 \\
Co-occurrence-HG & 99.650 & 70.665 & 70.070 & 31.734 & 35.632 & 34.712 & 98.204 & 71.074 & 70.722 \\
Static-HG      & 99.650 & 78.098 & 77.529 & 36.740 & 37.714 & 36.905 & 98.190 & 73.340 & 73.018 \\
\whg{} & 99.650 & \textbf{79.299} & \textbf{78.650} & 38.643 & 39.925 & 39.222 & 98.190 & 72.957 & 72.586 \\
\bottomrule
\end{tabular*}
\end{table*}

\begin{table}[t]
\caption{Paired-bootstrap support for the primary fixed-substrate key-space comparison. Differences are \whg{} minus \kgppr{} on seed-13 unique examples.}
\label{tab:bootstrap}
\centering
\small
\begin{tabular}{@{}lrr@{}}
\toprule
Dataset & $\Delta$F1 & 95\% CI \\
\midrule
2Wiki & $+3.426$ & $[+2.877,+3.984]$ \\
MuSiQue & $+3.592$ & $[+1.917,+5.155]$ \\
HotpotQA & $-0.689$ & $[-1.453,+0.071]$ \\
\bottomrule
\end{tabular}
\end{table}

\subsection{From Standalone Retrieval to Evidence Control}

To interpret the standalone boundary, we ask whether hypergraph tuples lack useful evidence or whether the system selects the wrong support from them. Table~\ref{tab:phase1_chain} gives a no-gold baseline, two oracle upper bounds, and train-to-dev variants. The oracle rows use support information and are excluded from bolding. Oracle answer-path selection raises F1 from 79.299 to 90.884 on 2WikiMultiHopQA, from 39.925 to 74.330 on MuSiQue, and from 72.957 to 91.111 on HotpotQA. Thus the tuple cache is not the hard ceiling; selecting and ranking answer-supporting evidence is the bottleneck. The gold-support construction oracle, which assembles hyperedges directly from gold passages, reaches 91.357, 77.305, and 93.275 F1. The gap between the two oracles is small on 2Wiki (+0.473 F1) and moderate on MuSiQue (+2.975 F1), leaving modest construction headroom on MuSiQue.

The train-to-dev group tests how much of this selection gap can be recovered without dev labels. The train-derived support scorer raises F1 to 87.212 on 2Wiki, 54.252 on MuSiQue, and 81.525 on HotpotQA, closing 68.3\%, 41.6\%, and 47.2\% of the oracle-selection gaps. Learned HKVM calibration then reaches 88.391, 56.754, and 82.929 F1. For 2Wiki and MuSiQue, calibration improves over the matched deterministic support-scorer control by +1.179 and +2.133 F1 with positive paired-bootstrap intervals; HotpotQA shows a descriptive +1.404 step before the final controller. These results motivate WHG-KV as an evidence-control signal, while leaving mechanism attribution to the controller and source-ablation analyses below. They do not show that structural proxy features alone caused the improvement.

\begin{table*}[t]
\caption{Evidence-selection diagnosis and controller validation. The oracle rows are diagnostic upper bounds and excluded from bolding; bold marks the best non-oracle result within each dataset.}
\label{tab:phase1_chain}
\centering
\small
\begin{tabular*}{1.0\textwidth}{@{\extracolsep{\fill}} l rrr rrr rrr @{}}
\toprule
& \multicolumn{3}{c}{\textbf{2WikiMultiHopQA}} & \multicolumn{3}{c}{\textbf{MuSiQue}} & \multicolumn{3}{c}{\textbf{HotpotQA}} \\
\cmidrule(lr){2-4} \cmidrule(lr){5-7} \cmidrule(lr){8-10}
Variant & F1 & EM & AR@10 & F1 & EM & AR@10 & F1 & EM & AR@10 \\
\midrule
\multicolumn{10}{@{}l}{\emph{No-gold structured baseline}} \\
No-gold WHG retrieval & 79.299 & 78.650 & 99.650 & 39.925 & 39.222 & 38.643 & 72.957 & 72.586 & 98.190 \\
\midrule
\multicolumn{10}{@{}l}{\emph{Oracle diagnostic upper bound}} \\
Oracle answer-path selection & 90.884 & 90.832 & 99.634 & 74.330 & 74.059 & 73.066 & 91.111 & 91.047 & 98.190 \\
Oracle gold-support construction & 91.357 & 91.333 & 99.634 & 77.305 & 77.120 & 71.163 & 93.275 & 93.248 & 98.177 \\
\midrule
\multicolumn{10}{@{}l}{\emph{Train-to-dev variants}} \\
Train-derived support scorer & 87.212 & 86.967 & 98.378 & 54.252 & 53.662 & 50.228 & 81.525 & 81.269 & 96.016 \\
Learned HKVM calibration & 88.391 & 88.181 & 99.030 & 56.754 & 56.213 & 53.303 & 82.929 & 82.705 & 96.439 \\
Dense-aware HKVM controller & \textbf{88.846} & \textbf{88.658} & \textbf{100.000} & \textbf{65.073} & \textbf{64.722} & \textbf{64.115} & \textbf{85.810} & \textbf{85.627} & \textbf{100.000} \\
\bottomrule
\end{tabular*}
\end{table*}

\subsubsection{Dense-Aware HKVM Controller}

To test complementarity rather than replacement, we treat WHG-KV as a control signal over dense candidates. Table~\ref{tab:phase1d_controller} reports raw controller outcomes. A 5,000-sample paired-bootstrap audit over frozen prediction files gives positive F1 intervals for all six controller-vs-comparator tests in the supplementary material; lower bounds are +10.740/+0.368 on 2Wiki, +5.791/+7.327 on MuSiQue, and +5.491/+2.555 on HotpotQA against ColBERTv2/learned HKVM. On 2WikiMultiHopQA, learned HKVM calibration is already strong at 88.391 F1, so the dense-aware controller adds a smaller gain over calibration (+0.455 F1) while preserving the large advantage over ColBERTv2 (+11.084 F1).

MuSiQue gives the clearest dense/HKVM combination case. ColBERTv2 reaches 58.309 F1, learned HKVM calibration reaches 56.754 F1, and source-only controllers remain near those references at 58.247 and 56.760 F1. Combining the two signals in the dense-aware HKVM controller raises F1 to 65.073, improving over ColBERTv2 by +6.763 and over learned HKVM calibration by +8.319.

HotpotQA follows the same train-to-dev protocol, with a different interpretation. ColBERTv2 reaches 79.844 F1, learned HKVM calibration reaches 82.929, and the dense-aware controller reaches 85.810. This improves over ColBERTv2 by +5.966 F1 and over learned HKVM calibration by +2.881 F1. It does not reverse the standalone HotpotQA finding in Table~\ref{tab:main_results}; it separates two roles for the same signal: standalone WHG-KV is not the strongest retriever, but its rank/score outputs help reorder dense candidates.

Across datasets, 2Wiki is HKVM-favorable, MuSiQue is the clearest dense/HKVM combination case, and HotpotQA is a controller-mediated recovery case rather than a standalone structured-retrieval win. Simple list-fusion baselines did not reproduce the controller gains (supplementary material), and source-only variants show that single-source re-ranking does not explain the combined gain.

Out-of-fold HKVM training inputs address the main leakage concern: each train example receives HKVM inputs from models trained on other folds, while dev evaluation uses frozen prediction files. Together with the source-only controls, this supports the narrower claim that dense and WHG-KV rank/score signals jointly carry useful passage-reordering information under this protocol.

\begin{table*}[t]
\caption{Dense-aware HKVM controller outcomes with out-of-fold HKVM training signals. Bold marks the best F1 and EM within each dataset; paired-bootstrap deltas for the final controller are reported in the supplementary material.}
\label{tab:phase1d_controller}
\centering
\small
\begin{tabular*}{1.0\textwidth}{@{\extracolsep{\fill}} l rrr rrr rrr @{}}
\toprule
& \multicolumn{3}{c}{\textbf{2WikiMultiHopQA}} & \multicolumn{3}{c}{\textbf{MuSiQue}} & \multicolumn{3}{c}{\textbf{HotpotQA}} \\
\cmidrule(lr){2-4} \cmidrule(lr){5-7} \cmidrule(lr){8-10}
Variant & F1 & EM & AR@10 & F1 & EM & AR@10 & F1 & EM & AR@10 \\
\midrule
\multicolumn{10}{@{}l}{\emph{Reference predictors}} \\
ColBERTv2 & 77.763 & 77.195 & 100.000 & 58.309 & 57.758 & 49.317 & 79.844 & 79.635 & 100.000 \\
Learned HKVM calibration & 88.391 & 88.181 & 99.030 & 56.754 & 56.213 & 53.303 & 82.929 & 82.705 & 96.439 \\
\midrule
\multicolumn{10}{@{}l}{\emph{Source-only controller controls}} \\
Controller dense-only & 77.679 & 77.107 & 100.000 & 58.247 & 57.758 & 52.172 & 79.909 & 79.671 & 100.000 \\
Controller HKVM-only & 88.415 & 88.205 & 100.000 & 56.760 & 56.213 & 58.640 & 82.929 & 82.705 & 100.000 \\
\midrule
\multicolumn{10}{@{}l}{\emph{Dense-aware controller}} \\
Dense-aware HKVM controller & \textbf{88.846} & \textbf{88.658} & 100.000 & \textbf{65.073} & \textbf{64.722} & \textbf{64.115} & \textbf{85.810} & \textbf{85.627} & 100.000 \\
\bottomrule
\end{tabular*}
\end{table*}

\subsection{The Gain Is WHG-KV-Specific, Not Generic Structured Stacking}

To rule out a generic second-source stacking explanation, we keep the controller protocol fixed and replace only the structured source. Table~\ref{tab:source_controller_ablation} tests this under the primary ColBERTv2 first-stage protocol: ColBERTv2 predictions, train-to-dev split, controller family, and hyperparameters are unchanged. Figure~\ref{fig:controller_evidence} visualizes the same source-effect trend with BM25 and Contriever diagnostics. On 2WikiMultiHopQA, Static-HG and Weighted-KG-PPR improve over ColBERTv2 by only +2.390 and +3.007 F1, while WHG-KV improves by +11.084. On MuSiQue, Co-occurrence-HG, Static-HG, and Weighted-KG-PPR remain below ColBERTv2 at 56.399, 55.258, and 56.853 F1, whereas ColBERTv2 + WHG-KV reaches 65.073 (+6.763). On HotpotQA, the best non-WHG source improves by +0.569 F1, while WHG-KV improves by +5.966.

For the three primary ColBERTv2 source-level WHG-KV rows, paired-bootstrap F1 intervals versus ColBERTv2 are fully positive: 2Wiki $[+10.722,+11.442]$, MuSiQue $[+5.813,+7.753]$, and HotpotQA $[+5.479,+6.456]$. WHG-KV is also above the best non-WHG source by +8.077, +8.219, and +5.397 F1. Table~\ref{tab:first_stage_robustness} reports BM25 and Contriever runs as absolute Base/Non/WHG F1 values because they test first-stage robustness and final combination strength, not a second full source-replacement table. Co-occurrence-HG is a matched Hyper-RAG-style source in our cached pipeline, not a reproduction of Hyper-RAG's original high-order extraction system.

\begin{table*}[t]
\caption{Primary ColBERTv2 source-level controller ablation. Each non-baseline row keeps the ColBERTv2 first-stage predictions and controller protocol fixed while swapping only the structured prediction source supplied to the controller; $\Delta$F1 is relative to ColBERTv2 only within the same dataset. Bold marks the best F1 and $\Delta$F1.}
\label{tab:source_controller_ablation}
\centering
\small
\begin{tabular*}{0.8\textwidth}{@{\extracolsep{\fill}} l rr rr rr @{}}
\toprule
& \multicolumn{2}{c}{\textbf{2WikiMultiHopQA}} & \multicolumn{2}{c}{\textbf{MuSiQue}} & \multicolumn{2}{c}{\textbf{HotpotQA}} \\
\cmidrule(lr){2-3} \cmidrule(lr){4-5} \cmidrule(lr){6-7}
Variant & F1 & $\Delta$F1 & F1 & $\Delta$F1 & F1 & $\Delta$F1 \\
\midrule
ColBERTv2 only & 77.763 & +0.000 & 58.309 & +0.000 & 79.844 & +0.000 \\
+ Co-occurrence-HG & 77.343 & -0.419 & 56.399 & -1.911 & 79.418 & -0.426 \\
+ Static-HG & 80.153 & +2.390 & 55.258 & -3.051 & 80.161 & +0.317 \\
+ Weighted-KG-PPR & 80.769 & +3.007 & 56.853 & -1.456 & 80.413 & +0.569 \\
+ WHG-KV & \textbf{88.846} & \textbf{+11.084} & \textbf{65.073} & \textbf{+6.763} & \textbf{85.810} & \textbf{+5.966} \\
\bottomrule
\end{tabular*}
\end{table*}

\subsubsection{First-stage source robustness}

\begin{figure*}[t]
\centering
\includegraphics[width=\textwidth]{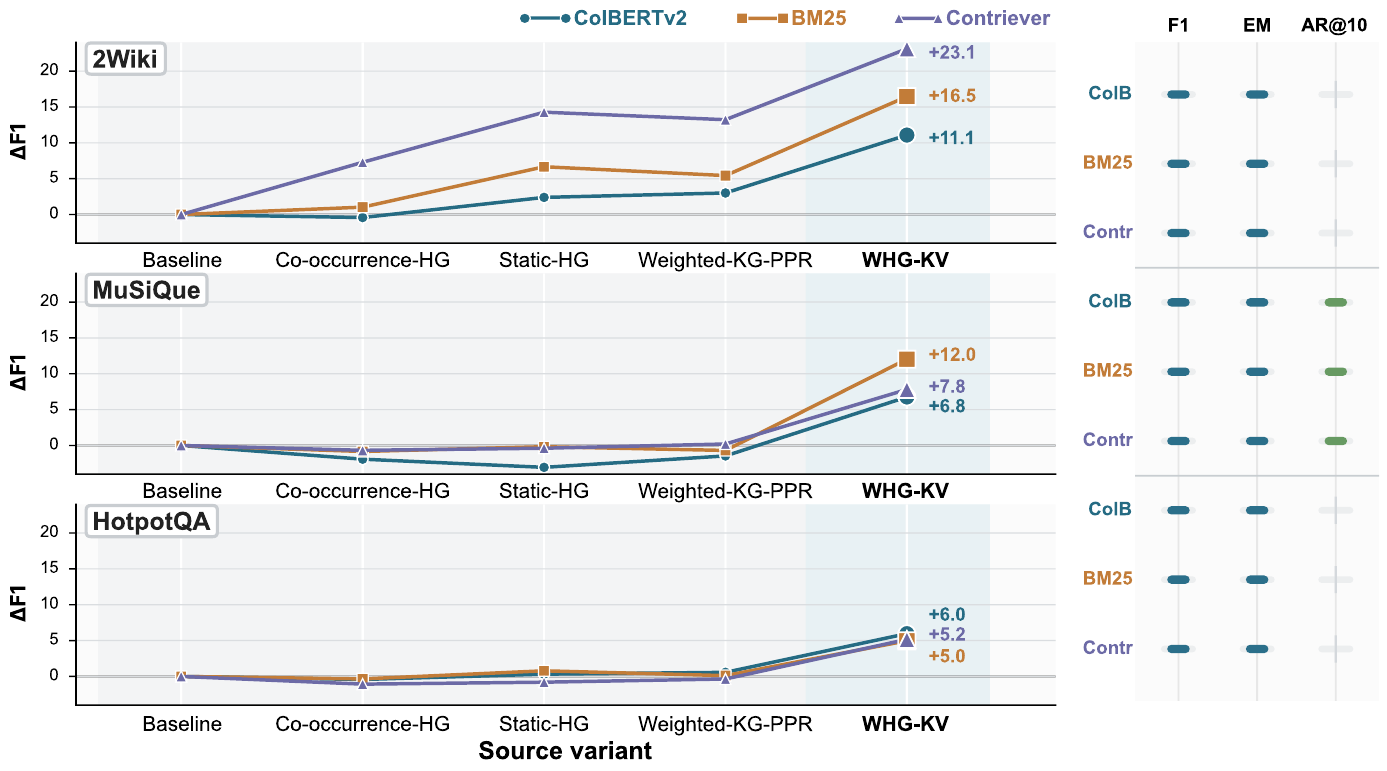}
\caption{Source-effect diagnostics for dense-aware evidence control. The left panels plot $\Delta$F1 relative to each dataset--first-stage baseline as the structured source is changed from no complement to Co-occurrence-HG, Static-HG, Weighted-KG-PPR, and WHG-KV. Colored curves denote ColBERTv2, BM25, and Contriever first-stage sources; endpoint labels mark the WHG-KV $\Delta$F1. The right metric strip summarizes WHG-KV margins over the best non-WHG complement for F1, EM, and AR@10. These results support a WHG-KV-specific source effect, with AR@10 margins interpreted under the recall ceilings shown in Table~\ref{tab:phase1d_controller}.}
\label{fig:controller_evidence}
\end{figure*}

To test whether the pattern is tied to late-interaction ColBERTv2, we repeat the source-level diagnostic with BM25 and Contriever. Table~\ref{tab:first_stage_robustness} reports absolute source-level F1 after replacing the first-stage source, while Figure~\ref{fig:controller_evidence} shows the corresponding $\Delta$F1 profiles. For each first-stage source, the non-WHG column selects the best complement among Co-occurrence-HG, Static-HG, and Weighted-KG-PPR.

With ColBERTv2, BM25, and Contriever as first-stage sources, WHG-KV is the strongest structured complement by point estimate in every row of Table~\ref{tab:first_stage_robustness}. Across the nine dataset--first-stage rows, WHG-KV improves F1 over the corresponding first-stage baseline by +4.971 to +23.119; all nine baseline-vs-WHG paired-bootstrap F1 intervals are fully positive, with lower bounds from +4.554 to +22.668. The absolute best WHG-KV pairing is dataset-dependent: ColBERTv2+WHG-KV is highest on 2WikiMultiHopQA and MuSiQue, while BM25+WHG-KV is highest on HotpotQA. These checks extend Table~\ref{tab:source_controller_ablation}; they are train-to-dev diagnostics, not hidden-test or full retrieval-backbone claims.

\begin{table*}[t]
\caption{First-stage source diagnostics for the source-level controller. Base is source-alone F1; Non is the best absolute F1 among Co-occurrence-HG, Static-HG, and Weighted-KG-PPR complements; WHG is the absolute F1 after adding WHG-KV. Bold marks the highest WHG value within each dataset.}
\label{tab:first_stage_robustness}
\centering
\small
\begin{tabular*}{1.0\textwidth}{@{\extracolsep{\fill}} l rrr rrr rrr @{}}
\toprule
& \multicolumn{3}{c}{\textbf{2WikiMultiHopQA}} & \multicolumn{3}{c}{\textbf{MuSiQue}} & \multicolumn{3}{c}{\textbf{HotpotQA}} \\
\cmidrule(lr){2-4} \cmidrule(lr){5-7} \cmidrule(lr){8-10}
First-stage source & Base & Non & WHG & Base & Non & WHG & Base & Non & WHG \\
\midrule
ColBERTv2  & 77.763 & 80.769 & \textbf{88.846} & 58.309 & 56.853 & \textbf{65.073} & 79.844 & 80.413 & 85.810 \\
BM25       & 71.873 & 78.539 & 88.325 & 46.045 & 45.872 & 58.083 & 81.835 & 82.618 & \textbf{86.806} \\
Contriever & 65.498 & 79.773 & 88.618 & 55.565 & 55.773 & 63.378 & 79.930 & 79.574 & 85.096 \\
\bottomrule
\end{tabular*}
\end{table*}

\begin{figure*}[t]
\centering
\includegraphics[width=\textwidth]{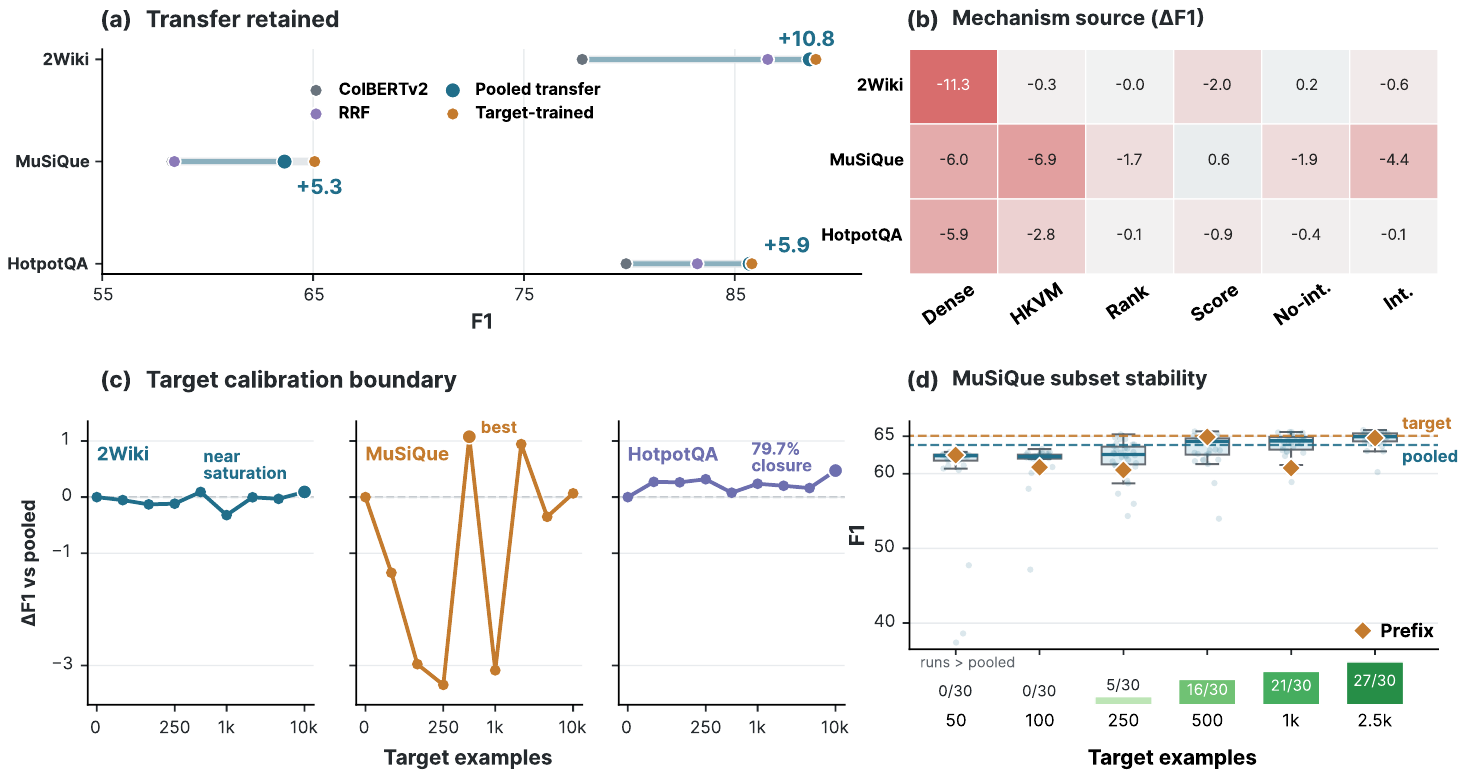}
\caption{Transfer and target-calibration boundary over frozen prediction artifacts. (a) Leave-one-target pooled transfer; RRF denotes reciprocal-rank fusion. (b) Feature ablation as $\Delta$F1 relative to full dense+HKVM transfer; Dense/HKVM/Rank/Score/No-int./Int. denote dense-only, HKVM-only, rank-only, score-only, no-interaction, and interaction-only variants. (c) Target calibration; size 0 is pooled transfer. (d) MuSiQue subset stability over 30 random subsets per size; diamonds mark prefix runs, dashed lines mark pooled/target-trained references, and the bottom strip counts runs above the seed-specific pooled baseline.}
\label{fig:transfer_calibration}
\end{figure*}

\subsection{Transfer and Target-Calibration Boundary}

The preceding controllers are target-specific. A four-part audit over frozen prediction artifacts asks whether the evidence-control signal transfers, what part transfers, and how much target-domain calibration remains useful. Figure~\ref{fig:transfer_calibration} summarizes the result; the supplementary material reports the full P0--P2b tables, bootstrap checks, and random-subset runs.

In the leave-one-target setting, the controller is trained on the two non-target benchmarks and evaluated on the held-out target. It reaches 88.555 F1 on 2WikiMultiHopQA, 63.645 on MuSiQue, and 85.706 on HotpotQA, exceeding both ColBERTv2 and reciprocal-rank fusion on all three targets. This is transfer evidence, not target-data independence: the target-trained controller remains strongest or near-strongest. The transfer ablation narrows what travels. Dense-only and HKVM-only variants are insufficient on MuSiQue and HotpotQA, whereas variants that keep dense and WHG-KV rank/score channels preserve most of the gain. Explicit dense-HKVM interaction features are not the driver: no-interaction and rank-only variants match or nearly match full transfer on 2Wiki and HotpotQA, and score-only slightly exceeds full by point estimate on MuSiQue. The supported mechanism is rank/score calibration across dense and WHG-KV channels, not handcrafted interaction terms or controller complexity.

Target-domain calibration gives the boundary. The pooled controller is already close to target-trained on 2Wiki. HotpotQA shows the cleanest calibration response, closing 53.7\% of the pooled-to-target-trained F1 gap at 250 target examples and 79.7\% at 10,000. MuSiQue is less orderly: its deterministic prefix curve reaches 64.914 F1 at 500 examples and 64.782 at 2,500, but drops at other prefix sizes. A repeated random-subset audit explains part of this behavior. At 2,500 target examples, 30 MuSiQue random-subset runs average 64.645 F1, close to the 65.073 target-trained controller, and 27/30 runs beat their seed-specific pooled-transfer baselines. Small subsets remain unstable, with 0/30 positive-delta runs at 50 and 100 examples. The transfer family supports evidence-control transfer while blocking stronger claims about monotonic sample efficiency, hidden-test generalization, or target-data-free deployment.

\section{Discussion and Limitations}

\textbf{Evidence organization as a controllable RAG layer.}
The experiments make the evidence-organization layer visible as a data-management object. Before the answer module reads a context, a system has already decided which candidate passages are retained, which tuples are extracted, which objects become retrieval keys, how key scores map back to passage values, and how prediction artifacts are audited. The standalone matrix tests the narrowest version of this idea: under a shared tuple cache, changing pairwise KG edges into answer-path hyperedges improves the structured-memory reference on 2WikiMultiHopQA and MuSiQue. The controller and transfer results broaden the point. Even when standalone structure is weaker than dense retrieval, WHG-KV exposes a reusable rank/score signal that helps decide which candidate passages form a better answer context.

\textbf{Complementarity rather than replacement.}
The results make little sense if HKVM is treated as a dense-retrieval replacement. MuSiQue is the clearest example: standalone WHG-KV is far below ColBERTv2, yet the dense-aware controller exceeds both source-only references. HotpotQA gives a different boundary: standalone WHG-KV improves support coverage over KG-PPR without improving answer F1, while the same signal helps reorder dense candidates. This is not a contradiction. It separates candidate discovery from evidence control. Lexical and dense retrievers are strong passage-value discovery mechanisms; HKVM contributes key-side organization that helps calibrate which candidate values should enter the final context.

\textbf{Mechanism attribution is intentionally narrow.}
The paper should not be read as evidence that explicit interaction features, high-order proxy features, or controller complexity alone caused the gains. Feature ablations show that explicit dense-HKVM interaction terms are not necessary for MuSiQue, and transfer ablations support the narrower interpretation that dense and WHG-KV rank/score channels jointly carry useful passage-reordering information. This negative evidence matters: it rules out the convenient story that a richer handcrafted feature set is the core contribution. The supported mechanism is key/value-separated evidence organization plus learned rank/score calibration, not a claim that every hypergraph-derived feature is causally important.

\textbf{Fixed substrate as control and boundary.}
The fixed-substrate protocol gives the study its interpretability and its main limit. It holds candidate passages, tuple caches, reader settings, and prediction artifacts fixed so key-space and controller changes can be interpreted. It also means that the results do not establish robustness to new extraction prompts, new LLM endpoints, new dense encoders, fresh online filtering, or different answer modules. \method{} depends on the quality, cost, and reproducibility of the extraction cache. The current system uses a fixed DeepSeek V4 Flash passage-level relation-triple cache with temperature 0.0; changing endpoint, prompt schema, high-order extraction policy, or extraction budget would create a new evidence matrix that should be reported with a new cache manifest rather than mixed into the current tables. The frozen passage, tuple, and prediction manifests make this boundary explicit and align with provenance principles that connect output claims to contributing data artifacts \cite{herschel2017provenance}.

\textbf{Baselines and related systems.}
Enhanced pairwise baselines with passage nodes and query-to-triple linking (supplementary material) do not close the WHG-KV gap on 2WikiMultiHopQA or MuSiQue, reducing the concern that the primary pairwise reference is too weak. HippoRAG 2 and EcphoryRAG remain closest related systems, but they are not direct numeric rows in the current matrix because the paper's controlled claim is key-space isolation rather than full-system performance ranking. HippoRAG 2 combines Llama-3.3-70B-Instruct, NV-Embed-v2, and online LLM triple filtering \cite{hipporag2}; EcphoryRAG uses Phi4/Ollama, bge-m3 embeddings, sampled evaluation, dataset-tuned final context sizes, and no exact public split/code path in our audit \cite{ecphoryrag}. Reproducing either system without aligned models, splits, prompts, and online filtering budgets would confound the structured-vs-KG comparison rather than isolate evidence organization.

\textbf{Parameter, cost, and generalization boundaries.}
The supplementary material reports diffusion-depth, pruning, hyperparameter, cost, and held-out-slice diagnostics. These are boundary checks, not alternate main systems. Changing traversal depth from $T=1$ to $T=2$ helps 2Wiki slightly, is neutral to negative on MuSiQue, and does not change the HotpotQA standalone conclusion. Hyperedge pruning improves HotpotQA's standalone WHG-KV point estimate at a 25\% keep point, but the result remains below BM25 and dense retrieval, so pruning is not a hidden tuned fix. The cost audit measures fixed-substrate scoring and controller replay over saved artifacts; it does not include fresh LLM extraction, full-corpus indexing, reader-model latency, online queues, or deployment throughput. Finally, the P0--P2b transfer family supports within-artifact transfer and calibration boundaries, not official hidden-test or cross-pipeline generalization.

\textbf{What this enables.}
The paper is more useful as a research direction than as a final universal retriever. Evidence organization can become a benchmark dimension in its own right: future systems can be evaluated by how they organize answer-supporting evidence into controllable contexts, not only by whether they retrieve relevant passages. HKVM points to learned hypergraph key construction, retriever-controller co-training, extraction-substrate robustness, provenance-aware evidence-organization benchmarks, and end-to-end models that learn key construction jointly with passage-value encoding. The present paper is deliberately narrower than those systems. Its contribution is controlled fixed-substrate evidence that the key/value separation is worth carrying forward.

\section{Conclusion}

This paper tested whether changing the retrieval key space from pairwise graph keys to answer-path hypergraph keys improves evidence organization for multi-hop RAG under a fixed extraction substrate. The answer is positive, but bounded. \whg{} outperforms \kgppr{} on 2WikiMultiHopQA ($+3.426$ F1) and MuSiQue ($+3.592$ F1), and enhanced pairwise baselines do not close this gap. This supports the key-space claim rather than a weak-baseline explanation. Standalone answer-path hypergraph retrieval is not a dense-retrieval replacement: HotpotQA exposes a case where better support coverage does not improve answer F1, and MuSiQue shows that structure alone remains far below ColBERTv2.

Beyond the standalone comparison, \whg{} acts as a reusable evidence-control signal. A lightweight controller combining frozen ColBERTv2 and WHG-KV features improves over ColBERTv2 by $+11.084$ F1 on 2WikiMultiHopQA, $+6.763$ on MuSiQue, and $+5.966$ on HotpotQA. Source-level and first-stage diagnostics support a WHG-KV-specific interpretation: matched non-WHG structured sources do not reproduce the gain, and the same pattern appears under BM25 and Contriever first-stage retrieval. Transfer and target-calibration audits further narrow the mechanism to dense/HKVM rank-score calibration under frozen prediction artifacts. The signal transfers, but target calibration and sample composition still matter.

The fixed-substrate protocol, source-level ablation design, transfer audit, and open artifact offer a starting point for this broader direction. Future work can extend the protocol to new extraction substrates, answer modules, and retriever families. The same key-value-separated framing can guide end-to-end models that learn hypergraph key construction jointly with passage-value encoding, and support standardized evidence-organization benchmarks beyond passage matching.

\section*{Acknowledgment}

AI-generated content disclosure: The authors used Codex Desktop for implementation debugging, data-processing checks, plotting-script scaffolds, LaTeX compilation and format diagnostics, citation-format checking, and English-language polishing. Figure source data, plotting scripts, captions, and manuscript claims were reviewed and edited by the authors, and all citations were retrieved and verified manually. The tool provided drafting and diagnostic support; it was not used for autonomous scientific decision-making, benchmark-label creation, hidden-result generation, or automatic final-claim selection. The authors remain responsible for the correctness and originality of the final submission.

\end{document}